\documentstyle[12pt]{article}
\def\be{\begin{equation}}
\def\ee{\end{equation}}
\def\ba{\begin{eqnarray}}
\def\ea{\end{eqnarray}}
\def\a{\alpha}
\def\t{\theta}

\begin{document}
\renewcommand{\theequation}{\arabic{equation}}
\newcommand{\beq}{\begin{equation}}
\newcommand{\eeq}[1]{\label{#1}\end{equation}}
\newcommand{\ber}{\begin{eqnarray}}
\newcommand{\eer}[1]{\label{#1}\end{eqnarray}}
\begin{titlepage}
\begin{center}
       August, 1992           \hfill    IASSNS-HEP-92/55\\
                              \hfill    PUPT-1337\\
                              \hfill    hep-th/9208076\\

\vskip .3in

{\large \bf On Cosmological String Backgrounds \\
            with Toroidal Isometries }
\vskip .3in

{\bf Amit Giveon} \footnotemark \\

\footnotetext{e-mail address: GIVEON@IASSNS.bitnet\\Address after
September 15: Racah Inst. of Physics, The Hebrew University, Jerusalem
91904, ISRAEL.\\ e-mail: GIVEON@HUJIVMS.bitnet}

\vskip .1in

{\em School of Natural Sciences, Institute for Advanced Study\\
  Olden Lane, Princeton, NJ 08540, USA} \\

\vskip .15in



{\bf Andrea Pasquinucci} \\ \vskip .1in

{\em Joseph Henry Laboratories, Department of Physics\\
     Princeton University, Princeton, NJ 08544, USA}\\

\vskip .1in

\end{center}

\vskip .2in

\begin{center} {\bf ABSTRACT } \end{center}
\begin{quotation}\noindent
A large class of cosmological solutions (of the Einstein equations)
in string theory, in the presence of
Maxwell fields, is obtained by $O(d,d)$ transformations of simple
backgrounds with $d$ toroidal isometries.
In all the examples in which  we find a
(closed) expanding universe, such that the
universe admits a smooth, complete initial value hypersurface,
a naked singularity may form only at the time when the universe collapses.
The discrete symmetry group $O(d,d,Z)$ identifies different
cosmological solutions with a background corresponding to a
(relatively) simple CFT, and therefore,
may be useful in understanding the properties of
naked singularities in string theory.

\end{quotation}
\end{titlepage}
\vfill
\eject
\def\baselinestretch{1.2}
\baselineskip 16 pt
To find realistic cosmological solutions in
string theory, one represents a classical
solution by $M\times K$, where $M$ is a $2-d$
conformal field theory (CFT)
with a four dimensional space-time, and with a
central charge $c=4$, and $K$ is some internal space
represented also by a CFT.
Moreover, though the possible formation of
singularities is one of the questions of interest, we do not want the
singularities to be ``built in'' in the initial conditions. We want $M$
to admit a smooth, complete initial value hypersurfaces.

In ref. \cite{NW}, Nappi and Witten presented a CFT that describes a
closed expanding universe in string theory. This
universe is a cosmological solution
with two toroidal isometries, namely, the background $M$ is independent of
two coordinates.\\

In this paper, we use the method of $O(d,d)$ transformations
\cite{V,GR} in order to
generate a large class of cosmological solutions in string theory.
In particular, the solutions of \cite{NW} are just a one parameter
sub-family, of the backgrounds obtained by $O(2,2)$ transformations of
(a suitable  analytic continuation of) a product of two $D=2$ black holes
\cite{W}. Moreover, when the internal space $K$ also has toroidal
isometries, one can turn on non-trivial gauge fields.
In the examples we discuss, we find an anisotropic expanding and
contracting closed universe, in the presence of Maxwell fields.
The topology of a spatial slice is that of a three sphere.

In the family of solutions we study, a naked singularity
can appear only at the initial time, when the volume of the universe
vanishes, and/or at the time when the universe recollapses.
The cosmic censorship is therefore protected, within a large class of
anisotropic cosmological solutions.\\


We begin by establishing our notation following \cite{GR}.
The two dimensional action describing a string propagating in
a (curved) background (of a metric, an
antisymmetric tensor, and a dilaton) in $D$-dimensions,
that is independent of $d$ coordinates is
\ba
S&=& \frac{1}{2\pi}\int d^2 z [\Sigma_{IJ}(x)\partial X^I \bar{\partial}
X^J-\frac{1}{4}\phi(x)R^{(2)}]
\label{S} \\
&=& \frac{1}{2\pi}\int d^2 z [E_{ij}(x)\partial \t^i \bar{\partial} \t^j
+ F_{ia}^2(x) \partial\t^i\bar{\partial}x^a + F_{ai}^1(x)\partial
x^a \bar{\partial}\t^i\nonumber \\
&&\qquad + F_{ab}(x)\partial x^a \bar{\partial} x^b -
\frac{1}{4}\phi(x)R^{(2)}]\ ,
\nonumber
\ea
where
\be
\{ X^I \}_{I=1...D}=\{\t^i,x^a\}_{i=1...d,\; a=d+1...D},\qquad
\t^i\equiv\t^i+2\pi,
\label{XI}
\ee
and
\be
\Sigma_{IJ}=G_{IJ}+B_{IJ}=\left(\matrix{E_{ij} & F_{ib}^2 \cr
                                    F_{aj}^1 & F_{ab} \cr}\right).
\label{Sigma}
\ee
$G_{IJ}$ and $B_{IJ}$ are the symmetric and antisymmetric parts of
$\Sigma_{IJ}$, respectively. The dilaton $\phi$ in (\ref{S}) is
coupled to the world-sheet curvature.

The background ($\Sigma(x),\phi(x)$) is independent of the $d$ coordinates
$\t^i$. If the background ($\Sigma(x), \phi(x)$) corresponds to a $2-d$
CFT, it can be transformed to new conformal backgrounds
by the action of a group isomorphic to  $O(d,d,R)$. The proof is given in
ref. \cite{GR}.\footnote{ ~The emphasize in ref. \cite{GR} is on the
discrete sub-group $O(d,d,Z)$, but the statement is true for
the full $O(d,d,R)$.}
The action of $O(d,d)$ on the background (\ref{Sigma})
(to leading order in the inverse string tension
$\a'$)\footnote{ ~For simplicity,
in (\ref{S}) and in the rest of this paper we set $\a'=\frac12$.
(For the insertion of $\a'$ factors see for example \cite{GSW}.)
}
was also described in
\cite{GR}. In this note we discuss examples for which $F^1=F^2=0$.
In this case only the block $E_{ij}$ in (\ref{Sigma})
transforms under $O(d,d)$, while the block
$F_{ab}$ is invariant. For the general case we refer the reader to
ref. \cite{GR}.

Next we describe the action of $O(d,d)$ on $E$ following \cite{GMR}.
The group $O(d,d)$ can be represented as $2d\times 2d$ dimensional
matrices $g$  preserving the bilinear form $J$:
\be
g=\left(\matrix{a & b\cr
                c & d\cr}\right),
\qquad J=\left(\matrix{0 & I\cr
                       I & 0\cr}\right),
\label{gJ}
\ee
where $a,b,c,d,I$ are $d\times d$ matrices, and
\be
g^t J g = J.
\ee
We define the action of $g$ on $E$ by fractional linear transformations:
\be
g(E)=E'=(aE+b)(cE+d)^{-1}.
\label{gE}
\ee
The dilaton transformation is
\be
g(\phi)=\phi'=\phi+\frac{1}{2}\log\left(
\frac{\det G_{ij}}{\det G'_{ij}}\right),
\label{gphi}
\ee
where $G_{ij}$ ($G'_{ij}$) is the symmetric part of $E_{ij}$ ($E'_{ij}$).

If ($E,\phi$) correspond to
(a leading order in $\a'$ of) a conformally exact
background, then ($E',\phi'$) are the leading order in $\a'$ of a
conformally exact background as well.
In particular, the low-energy effective action is invariant under $O(d,d)$
transformations~\cite{GMR,V}. In other words,
if ($G,B,\phi$) solve the $\beta=0$ equations to leading order, so
do ($G', B', \phi'$). One can therefore use the $O(d,d)$
transformations in order to generate new solutions to Einstein gravity
coupled to matter, starting with a known solution with $d$
isometries~\cite{V}.\\


We start to construct new solutions by exploiting the toroidal
isometries of the space-time background $M$. Consider the
$4-D$ line element
\be
d{\bf s}^2=-dt^2+ds^2+g(t)^2 d\t_1^2+f(s)^2 d\t_2^2.
\label{ds2}
\ee
We want this background to correspond to a CFT.\footnote{
{}~More precisely, we should start with
$d{\bf s}^2=k(-dt^2+g(t)^2d\t_1^2)+k'(ds^2+f(s)^2d\t_2^2)$,
and choose $k$ and $k'$ such that $c=4$.
To leading order in $\a'$ this condition is $k = k'$, and for
simplicity we take $k=k'=1$.}
To leading order in $\a'$ there are four nontrivial possibilities for
$g(t)$ and four possibilities for $f(s)$,
namely, $g(t)=\tan(t), \cot(t), \tanh(t)$ or $\coth(t)$,
and $f(s)=\tan(s), \cot(s), \tanh(s)$ or $\coth(s)$.\footnote{
{}~All these possibilities correspond to the exact CFT given by a
direct product of two cosets $\frac{G}{H}$, where $G$ is either $SU(2)$
or $SL(2)$ and $H$ is either $U(1)$ or ${\bf R}$. }
The dilaton is then given by
\be
\phi(s,t)=\phi_0+\log(\bar{f}(s)^2\bar{g}(t)^2),
\label{phits}
\ee
where $\phi_0$ is a constant and $\bar{g}(t)=\cos(t),\sin(t),\cosh(t)$ or
$\sinh(t)$,  $\bar{f}(s)=\cos(s), \sin(s), \cosh(s)$ or $\sinh(s)$,
respectively.

The background in (\ref{ds2}),(\ref{phits}) is a $D=4$ curved background
that is independent of $d=2$ coordinates. It has one time-like
coordinate $t$,
and three space-like coordinates. The background is time dependent, and
therefore, it describes a cosmological solution to Einstein equations.
The $2\times 2$ matrix $E$ is
\be
E(s,t)=\left(\matrix{g(t)^2 &0\cr
                     0&f(s)^2 \cr}\right).
\label{Efg}
\ee
We can now generate new cosmological solutions acting on ($E,\phi$)
in (\ref{Efg}), (\ref{phits})
with $O(2,2)$ transformations.

Next we discuss a particular one parameter sub-family
of rotations generating the solutions of \cite{NW}.
By transforming $E$ and $\phi$ with the group element $g_b\in O(2,2)$
(where $b$ is an arbitrary real number):
\be
g_b=\left(\matrix{0&I\cr
                  I&0\cr}\right)\left(\matrix{I&b\Theta\cr
                                              0&I\cr}\right),
\label{gb}
\ee
where
\be
I=\left(\matrix{1&0\cr
                0&1\cr}\right),\qquad
\Theta=\left(\matrix{0&1\cr
                    -1&0\cr}\right),
\label{IT}
\ee
one finds
that the new background matrix, $E'_b$, is given by adding a constant
antisymmetric background $\left(\matrix{0&b\cr
                                       -b&0\cr}\right)$ to $E$, and then
using duality \cite{GRVSW} to invert the background matrix,
namely,
\be
g_b(E)=E'_b=(E+b\Theta)^{-1}=
\frac{1}{f(s)^2 g(t)^2+b^2}\left(\matrix{f(s)^2 & -b\cr
                                             b &g(t)^2 \cr}\right).
\label{NW}
\ee
{}From eq. (\ref{gphi}) one gets for the new dilaton
\be
\phi^\prime (s,t) = \phi_0 + \log \left[\bar{f}(s)^2\bar{g}(t)^2
\left(f(s)^2g(t)^2 + b^2\right)\right]\ .
\label{NWphi}
\ee

When
\ba
f(s)=\tan(s)\ &,&\qquad  g(t)=\cot(t)\ ,\nonumber \\
\bar{f}(s)=\cos(s)\ &,& \qquad \bar{g}(t)=\sin(t)\ ,
\label{fgtan}
\ea
we recover the solution discussed by Nappi and Witten in
\cite{NW}.\footnote{ ~To compare with \cite{NW} one should use
$\sin\a=\frac{1-b^2}{1+b^2}$,  reintroduce $k$ and $k'$, and take
$k=k'$ very large such that the maximal size of the universe is of
order $k$, and the central
charge of $M$ is $c=4$ (to leading order in $1/k$).}
In this case the conformal background corresponds to a
closed expanding universe.\\


We will now consider the case in which the internal space $K$
also has toroidal isometries, and by rotating on non-trivial
elements of the background matrix interpolating between the space-time
$M$ and the internal space $K$, we will turn on the gauge fields.

The action  (\ref{S}) can be
generalized adding internal degrees of freedom, {\it i.e.} extra
compactified coordinates $Y^A$. (The coordinates $Y^A$ can be regarded as
either the internal coordinates of a
bosonic string or the extra coordinates of
a heterotic string).
Following ref. \cite{GR}, we start from the action
\be
S=\frac{1}{2\pi}\int d^2 z [\Sigma_{IJ}(x)\partial X^I \bar{\partial}
X^J + E_{AB}\partial Y^A \bar\partial Y^B - \frac{1}{4}\phi(x)R^{(2)}],
\label{SS}
\ee
where  the space-time coordinates
$X^I$ are described in (\ref{XI}),
the blocks structure of $\Sigma_{IJ}$ is given in (\ref{Sigma}),
and  $E_{AB}$ is a constant matrix,  $A,B=1, \dots, d_{int}$.
The background fields $\Sigma_{IJ}(x), \phi(x),
E_{AB}$ are independent of the coordinates ($\theta^i, Y^A$).

We choose the initial structure of the background as before,
namely, the block
$E_{ij}$ is again given by eq. (\ref{Efg}), $F^1=F^2=0$,
and $F=\left(\matrix{-1 & 0\cr 0 & 1 \cr}\right)$.
It is now possible to make a more
general rotation using the group $O(2,2+d_{int})\subset
O(2+d_{int},2+d_{int})$ acting by the fractional
linear transformations (\ref{gE})
on the $(2+d_{int}) \times (2+d_{int})$ matrix
\be
\Xi = \left( \matrix{ E_{ij} & 0 \cr 0 & E_{AB}\cr}\right),
\ee
such that
\be
\Xi^\prime = \left( \matrix{ E_{ij}^\prime +
A_{iA}(G^{-1})^{AB} A_{jB} & 2A_{iA} \cr 0 & E_{AB}\cr}\right),
\label{Xip}
\ee
where $(G^{-1})^{AB}$ is the inverse of $G_{AB}$ defined by $E_{AB}=
G_{AB} + B_{AB}$.
The structure of $\Xi^\prime$ is such that $E'_{ij}=G'_{ij}+B'_{ij}$
gives the
correct metric after a dimensional reduction from $4+d_{int}$
dimensions to
four space-time dimensions. The structure of the
second line in the matrix $\Xi^\prime$
allows the internal coordinates to be the extra coordinates of a
heterotic string.

The new background in (\ref{Xip})  corresponds to the following
(curved) bosonic or heterotic two dimensional action (neglecting
worldsheet fermions in the heterotic case):
\be
S^\prime =\frac{1}{2\pi}\int d^2 z [\Sigma_{IJ}^\prime(x)
\partial X^I \bar{\partial} X^J + 2A_{IA} (x) \partial X^I \bar\partial
Y^A + E_{AB}\partial Y^A \bar\partial Y^B -
\frac{1}{4}\phi^\prime (x) R^{(2)}]\ .
\label{SSS}
\ee
Both the starting background and the new metric,
antisymmetric tensor, dilaton and
gauge fields are
solutions of the equations of motion derived from
the effective action (see \cite{GSW} for a review)
\be
S_{eff}=\int d^4 X \sqrt{-G} e^{\phi} \left[
R^{(4)} + (\nabla \phi)^2 -
\frac1{12} H^2 - \frac14 {\rm Tr} F^2
\right]\ .
\label{Seffprime}
\ee
(The starting background has $H_{\mu\nu\rho}=F_{\mu\nu}=0$.)


For simplicity,
we now consider the case in which $d_{int}=1$, {\it i.e.} we
start from a five dimensional background where the fifth coordinate is
compact and by an $O(2,3)\subset O(3,3)$
rotation we will introduce a Maxwell field $A_i$. (The discussion can be
easily generalized to the internal dimension
$d_{int}$ that is needed for criticality.)

We fix $E_{AB} = 1$ and we rotate $\Xi$ by the $g_{a,b,c}\in O(2,3)\subset
O(3,3)$ matrix
\be
g_{a,b,c} = \left(\matrix{0 & I\cr I & 0 \cr}\right)
\left(\matrix{I & \Theta\cr 0 & I \cr}\right)
\left(\matrix{A & 0\cr 0 & (A^t)^{-1} \cr}\right)
\label{gabc}
\ee
where
\be
\Theta = \left(\matrix{0 & b & c \cr -b & 0 & a \cr -c & -a & 0 }
\right),  \qquad\quad
A = \left(\matrix{1 & 0 & c \cr 0 & 1 & a \cr 0 & 0 & 1 }\right)\ ,
\ee
$I$ is the $3\times 3$ identity matrix and $a,b,c$ are arbitrary
real numbers.

After the rotation, the non-zero components of the metric
$G^\prime_{IJ}$~\footnote{ ~Recall that in eq. (\ref{SSS})
$\Sigma^\prime_{IJ} = G^\prime_{IJ} + B^\prime_{IJ}$.},
antisymmetric tensor $B^\prime_{IJ}$, gauge field  $A_I$ and
dilaton $\phi^\prime$, are:
\ba
&& G^\prime_{tt} = -1 \qquad\qquad\qquad\qquad\qquad \ \
G^\prime_{ss} = 1 \nonumber\\
&& G^\prime_{\theta_1\theta_1} = \frac1{\Delta^2} \left[ g(t)^2 \left(
a^2 + f(s)^2\right)^2 + f(s)^2 \left(ac + b \right)^2 \right]
\nonumber\\
&& G^\prime_{\theta_1\theta_2} = \frac1{\Delta^2} \left[ b\left(
a^2g(t)^2 - c^2 f(s)^2\right) - ac \left( a^2 g(t)^2 + c^2 f(s)^2 +
2 f(s)^2 g(t)^2\right) \right]
\nonumber\\
&& G^\prime_{\theta_2\theta_2} = \frac1{\Delta^2} \left[ f(s)^2 \left(
c^2 + g(t)^2\right)^2 + g(t)^2 \left(ac - b\right)^2 \right]
\nonumber\\
&& A_{\theta_1} = \frac1{\Delta} \left[ ab - cf(s)^2\right]
\qquad \qquad \  A_{\theta_2} = \frac1{\Delta}
\left[ -bc - a g(t)^2\right] \nonumber\\
&& B^\prime_{\theta_1\theta_2} = -\frac{b}{\Delta} \qquad\qquad
\qquad\qquad\qquad
\phi^\prime = \phi_0 + \log \left[ \bar{f}(s)^2 \bar{g}(t)^2
\,\cdot\, \Delta\right]\nonumber\\
&& \det\left[ - G^\prime\right] = \frac{g(t)^2 f(s)^2}{\Delta^2}
\label{GBA}
\ea
where
\be
\Delta = b^2 + a^2 g(t)^2 + c^2 f(s)^2 + g(t)^2 f(s)^2 \ .
\ee
Obviously, for $a=c=0$, one gets back
the space-time solutions in (\ref{NW}) and (\ref{NWphi}).\\

It is also possible to make a Weyl rescaling to bring the effective
action  (\ref{Seffprime}) to the Einstein form
\be
S_{eff}^E =\int d^4 X \sqrt{-G^E} \left[
R^{E\, (4)} - \frac12 (\nabla \phi)^2 -
\frac1{12} e^{2\phi}
H^2 - \frac14 e^{\phi} {\rm Tr} F^2 \right]\
\label{SeffE}
\ee
where the metric $G^E_{\mu\nu} = e^{\phi} G_{\mu\nu}$ is used to contract
the indices.\\

Notice that when $f$ and $g$ are given by eq.
(\ref{fgtan}) (for example) one has
\ba
\phi^\prime &=& \phi_0 + \log \left[ b^2 \sin^2(t) \cos^2(s) + a^2
\cos^2(t) \cos^2(s)\right. \\
&&\qquad\qquad + \left. c^2 \sin^2(t)\sin^2(s) + \cos^2(t)\sin^2(s)
\right]\ . \nonumber
\ea
For non-vanishing parameters $a,b$ and $c$, the dilaton is finite for
any $s,t$ (and therefore, Weyl rescaling to the Einstein form does not
change the properties of the universe described above).
For $a=0$ the dilaton diverges at ($t=0,\, s=0$), for $c=0$ it diverges
at ($t=\pi/2 , \, s=\pi/2$) and for $b=0$ it diverges at
($t=\pi/2,\,s=0$) ($t , s \in [0,\pi/2]$).
These divergences appear when a curvature singularity is formed.\\


We will now discuss  briefly the target space
properties of some of the models just presented.
{}From eq.(\ref{GBA}), we see that the universe described by that
class of solutions has zero volume
at $t=0$ (this is because $g(t=0)=0$ or $\infty$ for all possible
$g$'s). Depending on the choices of the functions $f, g$ and the
parameters $a,b,c$, the universe can then be an expanding and
contracting closed universe, in some cases collapsing at the end, or
an expanding open universe. (We will not
consider the models which do not have a smooth, complete initial value
hypersurface. An example of models with ``built in"
singularities in the initial conditions is given when $a=b=0$, $c \neq
0$ and the functions $f$ and $g$ given by eq. (\ref{fgtan}).)


Consider again the case when the functions $f$ and $g$
are given by eq. (\ref{fgtan}) (with $b$ and $c$ not both zero).
For this example, the backgrounds in (\ref{GBA})
describe an  anisotropic expanding and contracting universe, in the
presence of an antisymmetric tensor, a Maxwell  field, and a
dilaton. The
topology of a spatial slice is $S^3$, the three sphere.
The universe
recollapses at the time $t=\frac{\pi}{2}$.

When $b\neq 0$, this universe is  a modification of
the Nappi-Witten solution
\cite{NW} in the presence of the Maxwell  field
$(A_{\t_1}(s,t),A_{\t_2}(s,t))$ given in (\ref{GBA}).
The universe expands and collapses in the
presence of electro-magnetic currents. (There is no electric charge).

The case $b=0$ ($c\neq 0$) is particularly interesting.
In this case we find a classical solution
to dilaton gravity with a Maxwell field (given by the action
(\ref{Seffprime})
with $H=0$, namely, without the presence of the
antisymmetric background that vanishes for $b=0$).
This solution still corresponds to  a closed expanding universe as
described above.\\

Next we discuss the formation of singularities in the universe described
above. There are different cases depending on the parameters $a,b$,
and $c$.
For $a=0$ the universe is singular at the `big bang',
namely, when $t=0$ a
singularity is formed at $s=0$. For $b=0$ ($c=0$),
a naked singularity is about to form at $s=0$ ($s=\pi/2$) at the time
$t=\pi/2$, when the universe recollapses.
(For non-vanishing parameters $a,b$
and $c$ there are no singularities). Therefore, a naked singularity may
form only at the times when the universe collapses. \\

Finally, we point out briefly some consequences of the generalized
target space duality.
It was proven in ref. \cite{GR} that the elements of $O(d,d,Z)$ are
discrete symmetries of the space of curved string backgrounds that are
independent of $d$ coordinates. These discrete symmetries relate vacua
with geometries that in general are radically different, but that
correspond to the same $2-d$ conformal field theory (CFT).

In the examples we discussed, the matrix  $g_b$ in eq. (\ref{gb})
($g_{a,b,c}$ in eq. (\ref{gabc})) is in $O(2,2,Z)$ ($O(3,3,Z)$)
if $b\in Z$ ($a,b,c\in Z$).
For example, as a consequence,
all  the Nappi-Witten solutions \cite{NW} with
$\sin\a=\frac{1-b^2}{1+b^2}, b\in Z$ are equivalent as CFTs,
and in particular, they are all equivalent to the $b=0$ case, for
which the background is a direct product of (a suitable
analytic continuation) of
two $D=2$ black hole  solutions \cite{W}.
Moreover, all the solutions (\ref{GBA}) with $a,b,c\in Z$ are also
equivalent CFTs. Therefore, these
closed expanding universes in the
presence of electro-magnetic currents, are also equivalent as CFTs to the
direct product of (a suitable analytic continuation) of
two $D=2$ black hole  solutions, and an extra coordinate
compactified on a circle.

The discrete symmetries identify different cosmological solutions with a
background corresponding to a (relatively) simple CFT. Therefore, target
space dualities may be useful in understanding the properties of naked
singularities in string theory.\footnote{
This was pointed out for stringy black hole
singularities in \cite{GDVV}.}

\vskip .2in \noindent
{\bf Acknowledgements} \vskip .05in \noindent
We thank K. Bardakci, C. Nappi, and E. Witten for
useful discussions.
We would like to acknowledge the hospitality of the Theoretical Physics
Group at Lawrence Berkeley Laboratory where part of this work was done.
The work of A.G. is supported in part by DOE grant no.
DE--FG02--90ER40542. The research of A.P. is supported by an INFN
fellowship and partially by the NSF grant PHY90-21984.

\end{document}